\documentclass[conference]{IEEEtran}
\IEEEoverridecommandlockouts
\usepackage{cite}
\usepackage{amsmath,amssymb,amsfonts}
\usepackage{algorithmic}
\usepackage{graphicx}
\usepackage{textcomp}
\usepackage{xcolor}
\usepackage{float}
\def\BibTeX{{\rm B\kern-.05em{\sc i\kern-.025em b}\kern-.08em
    T\kern-.1667em\lower.7ex\hbox{E}\kern-.125emX}}
\begin{document}

\title{Acceleration of Atomistic NEGF: Algorithms, Parallelization, and Machine Learning\\
\thanks{This work was supported by the Swiss National Science Foundation (SNSF) under grant $\mathrm{n^\circ}$ 209358 (QuaTrEx). We acknowledge support from CSCS under projects lp16 and lp82. This research used resources of the Oak Ridge Leadership Computing Facility at the Oak Ridge National Laboratory, which is supported by the Office of Science of the U.S. Department of Energy under Contract No. DE-AC05-00OR22725 (project NEL107).}
}

\author{\IEEEauthorblockN{1\textsuperscript{st} Mathieu Luisier}
\IEEEauthorblockA{\textit{Integrated Systems Laboratory} \\
\textit{ETH Zurich}\\
Zurich, Switzerland \\
mluisier@iis.ee.ethz.ch}
\and
\IEEEauthorblockN{2\textsuperscript{nd} Nicolas Vetsch}
\IEEEauthorblockA{\textit{Integrated Systems Laboratory} \\
\textit{ETH Zurich}\\
Zurich, Switzerland \\
vetschn@iis.ee.ethz.ch}
\and
\IEEEauthorblockN{3\textsuperscript{rd} Alexander Maeder}
\IEEEauthorblockA{\textit{Integrated Systems Laboratory} \\
\textit{ETH Zurich}\\
Zurich, Switzerland \\
almaeder@iis.ee.ethz.ch}
\and
\IEEEauthorblockN{4\textsuperscript{th} Vincent Maillou}
\IEEEauthorblockA{\textit{Integrated Systems Laboratory} \\
\textit{ETH Zurich}\\
Zurich, Switzerland \\
vmaillou@iis.ee.ethz.ch}
\and
\IEEEauthorblockN{5\textsuperscript{th} Anders Winka}
\IEEEauthorblockA{\textit{Integrated Systems Laboratory} \\
\textit{ETH Zurich}\\
Zurich, Switzerland \\
awinka@iis.ee.ethz.ch}
\and
\IEEEauthorblockN{6\textsuperscript{th} Leonard Deuschle}
\IEEEauthorblockA{\textit{Integrated Systems Laboratory} \\
\textit{ETH Zurich}\\
Zurich, Switzerland \\
dleonard@iis.ee.ethz.ch}
\and
\IEEEauthorblockN{7\textsuperscript{th} Chen Hao Xia}
\IEEEauthorblockA{\textit{Integrated Systems Laboratory} \\
\textit{ETH Zurich}\\
Zurich, Switzerland \\
chexia@iis.ee.ethz.ch}
\and
\IEEEauthorblockN{8\textsuperscript{th} Manasa Kaniselvan}
\IEEEauthorblockA{\textit{Integrated Systems Laboratory} \\
\textit{ETH Zurich}\\
Zurich, Switzerland \\
mkaniselvan@iis.ee.ethz.ch}
\and
\IEEEauthorblockN{9\textsuperscript{th} Marko Mladenovi\'c}
\IEEEauthorblockA{\textit{Integrated Systems Laboratory} \\
\textit{ETH Zurich}\\
Zurich, Switzerland \\
mmladenovic@iis.ee.ethz.ch}
\and
\IEEEauthorblockN{10\textsuperscript{th} Jiang Cao}
\IEEEauthorblockA{\textit{Integrated Systems Laboratory} \\
\textit{ETH Zurich}\\
Zurich, Switzerland \\
jiang.cao@iis.ee.ethz.ch}
\and
\IEEEauthorblockN{11\textsuperscript{th} Alexandros Nikolaos Ziogas}
\IEEEauthorblockA{\textit{Integrated Systems Laboratory} \\
\textit{ETH Zurich}\\
Zurich, Switzerland \\
alziogas@iis.ee.ethz.ch}
}

\maketitle

\begin{abstract}
The Non-equilibrium Green’s function (NEGF) formalism is a particularly powerful method to simulate the quantum transport properties of nanoscale devices such as transistors, photo-diodes, or memory cells, in the ballistic limit of transport or in the presence of various scattering sources such as electron-phonon, electron-photon, or even electron-electron interactions. The inclusion of all these mechanisms has been first demonstrated in small systems, composed of a few atoms, before being scaled up to larger structures made of thousands of atoms. Also, the accuracy of the models has kept improving, from empirical to fully \textit{ab-initio} ones, e.g., density functional theory (DFT). This paper summarizes key (algorithmic) achievements that have allowed us to bring DFT+NEGF simulations closer to the dimensions and functionality of realistic systems. The possibility of leveraging graph neural networks and machine learning to speed up \textit{ab-initio} device simulations is discussed as well.

\end{abstract}


\section{Introduction}
The continuous shrinking of transistors' dimensions is posing significant design and fabrication challenges to the semiconductor industry. At the same time, it is bringing the featured sizes of nanoscale devices closer to what \textit{ab-initio} quantum transport (QT) simulators can handle within the framework of the atomistic non-equilibrium Green's function (NEGF) formalism, i.e., a few thousands \cite{deuschle}. To reach this number, several progresses have been made since the pioneering works of Taylor et al. \cite{taylor} and Brandbyge et al., combining density-functional theory (DFT) and NEGF at the beginning of the 2000s. Efficient numerical algorithms have been developed \cite{petersen}, the parallelization of the workload has been widely popularized \cite{klimeck}, collaborations with high-performance computing engineers have been established \cite{darve}, while the emergence of efficient hardware, in particular graphics processing units (GPUs), has paved the way for computer-aided investigations of complex phenomena in nano-transistors \cite{ziogas}.

Despite impressive evolution, most \textit{ab-initio} QT simulations are still restricted to the ballistic limit of transport, while electron-phonon or electron-electron interactions might play a detrimental role in nanoscale devices and should therefore be accounted for. Here, we give a brief overview of how the current limitations of QT tools can be addressed, present our implementation of the required models and algorithms into a novel, open source package called \textit{QuaTrEx}, and illustrate it with one example, the simulation of a silicon nano-ribbon in the presence of carrier-carrier scattering. We also suggest how machine learning can be used to bypass the generation of Hamiltonian matrices from first-principles.

\section{Ab-initio Quantum Transport}
Our approach relies on the combination of DFT with NEGF to shed light on the characteristics of nanostructures driven out of equilibrium by an external perturbation (voltage, thermal gradient, or optical signal). The first step consists of creating the Hamiltonian matrix of the system of interest with DFT. Packages relying on a localized basis set, such as CP2K \cite{cp2k} or transformations of plane-wave data into maximally localized Wannier functions (MLWFs) \cite{wannier}, can be utilized for that purpose. The dynamical, bare Coulomb, or electron-phonon coupling matrices can be constructed similarly.

\begin{figure}[!t]
\center
\includegraphics[width=\linewidth]{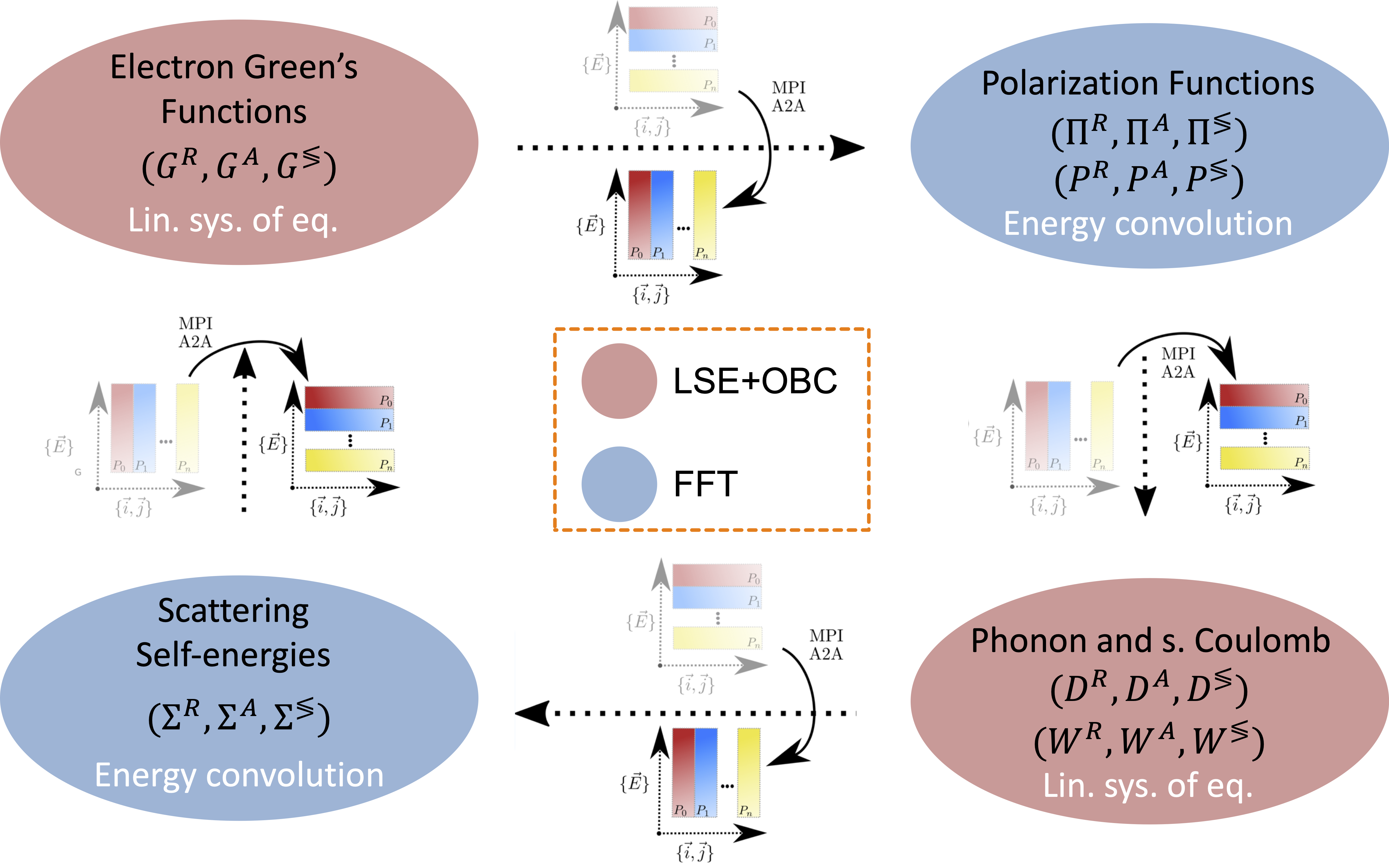}
\caption{Iterative scheme to compute the non-equilibrium Green's functions (NEGF) for electrons (retarded: $\mathbf{G}^R$) and lesser/greater: $\mathbf{G}^{\lessgtr}$ in the presence of electron-phonon and/or electron-electron interactions. To determine the corresponding scattering self-energies ($\mathbf{\Sigma}^{R,\lessgtr}$), the phonon ($\mathbf{D}^{R,\lessgtr}$) and screened Coulomb ($\mathbf{W}^{R,\lessgtr}$) Green's functions must be evaluated, which requires the knowledge of the phonon self-energy ($\mathbf{\Pi}^{R,\lessgtr}$) and/or of the polarization function ($\mathbf{P}^{R,\lessgtr}$). All these quantities are matrices $\mathbf{M}$ that depend on the energy ($E$) or frequency ($\omega$). Their entries ($M_{ij}$) represent correlations between two points located at $R_i$ and $R_{j}$. The calculation of all these matrices involves solving linear systems of equations and open boundary conditions (LSE+OBC, red blocks) or performing energy convolutions through fast Fourier transforms (FFT, blue blocks). In a parallel implementation, it is convenient to store all $M_{ij}$ entries for one or a few $E$/$\omega$ points when dealing with LSE+OBC, while it is preferable to have access to a few $M_{ij}$ for multiple $E$/$\omega$ when computing the $\mathbf{\Sigma}^{R,\lessgtr}$, $\mathbf{\Pi}^{R,\lessgtr}$, or $\mathbf{P}^{R,\lessgtr}$ terms with FFT. Data can be ``transposed'' with all-to-all communication operations from one representation ($E$,$\omega$;$i,j$) to the other ($i,j$;$E$,$\omega$).}
\label{fig2}
\end{figure}

Regardless of the particles (electron, phonon, photon, screened Coulomb) and interactions (electron-phonon, electron-photon, phonon-phonon, electron-electron) considered, the NEGF equations can be written in the following form
\begin{equation}
\mathbf{C}^{R}(\mathcal{E}) = (\mathbf{A}(\mathcal{E})-\mathbf{B}^R(\mathcal{E}))^{-1},
\label{eq:R}
\end{equation}
\begin{equation}
\left[\mathbf{C}^{R}(\mathcal{E})\right]^{-1}\cdot\mathbf{C}^{\gtrless}(\mathcal{E})\cdot \left[\left(\mathbf{C}^{R}(\mathcal{E})\right)^{H}\right]^{-1} =  \mathbf{B}^{\gtrless}(\mathcal{E}),
\label{eq:GL}
\end{equation}
\begin{equation}
\mathbf{B}(\mathcal{E})\propto\int d\mathcal{E}'\left(\mathbf{M_1}\cdot \mathbf{C}_1(\mathcal{E}-\mathcal{E}')\cdot \mathbf{M_2}\right) \odot \mathbf{C}_2(\mathcal{E}'),
\label{eq:S}
\end{equation}
where $\mathcal{E}\in\{E,\omega\}$ is the energy ($E$) or frequency ($\omega$) of the Green's function matrix $\mathbf{C}(\mathcal{E})$ and ``self-energy'' $\mathbf{B}$, which can be of retarded ($R$), lesser ($<$), or greater ($>$) type. Note that the momentum dependence is omitted for simplicity. The $\mathbf{C}$, $\mathbf{C_1}$, and $\mathbf{C_2}$ matrices are either equal to $\mathbf{G}$ (electron), $\mathbf{D}$ (phonon/photon), or $\mathbf{W}$ (screened Coulomb interaction), while $\mathbf{B}$ refers to the interactions between different physical systems and corresponds to either $\mathbf{P}$ (polarization), $\mathbf{\Sigma}$ (electron scattering self-energy), or $\mathbf{\Pi}$ (phonon/photon scattering self-energy). It also contains an open boundary condition term not shown here. The matrix $\mathbf{A}$ is made of the operator describing the system's intrinsic properties, i.e., of the Hamiltonian, dynamical, or bare Coulomb matrices. Finally, the $\mathbf{M_1}$ and $\mathbf{M_2}$ matrices represent the coupling between different populations (they are equal to the identity matrix in some cases), whereas $\odot$ indicates an element-wise or standard matrix-matrix multiplication. 

\begin{figure}[!t]
\center
\includegraphics[width=\linewidth]{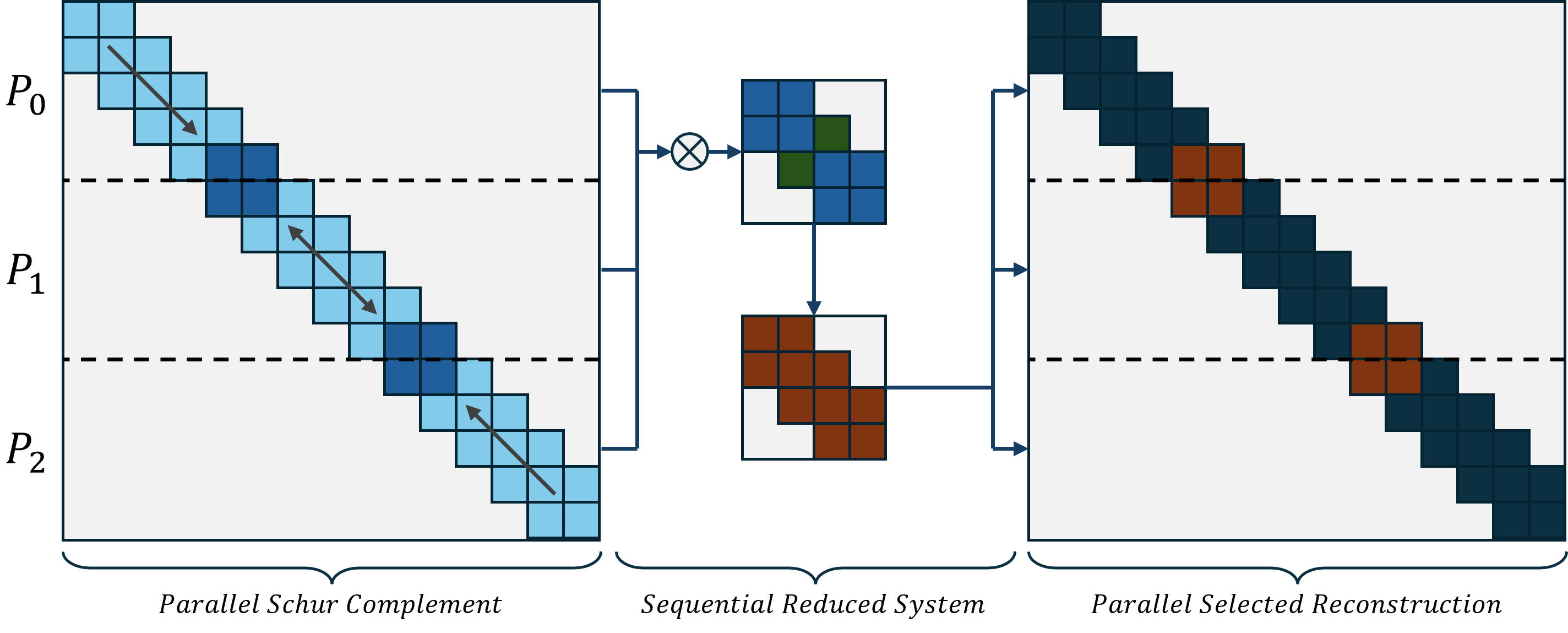}
\caption{Simplified representation of the Serinv algorithm \cite{serinv} for the parallel calculation of the NEGF equations (retarded and lesser/greater). It relies on the repeated application of Schur complements. Starting from a block tri-diagonal input matrix, $P$ partitions are created, each of them being attributed to different computing units (CPUs or GPUs). After local processing, a reduced system of equations arises, which must be solved sequentially before reconstructing selected entries of the Green's functions locally. Note the mapping between the large and reduced systems (blocks with the same color).}
\label{fig3}
\end{figure}

Typically, a block tri-diagonal (BT) shape is imposed on all $\mathbf{A}$ and $\mathbf{B}$ matrices by capping the maximum distance over which two particles can interact. As a consequence, only selected entries of $\mathbf{C}$ are needed, those corresponding to the sparsity pattern of $\mathbf{A}$ and $\mathbf{B}$ \cite{deuschle}. From a numerical point of view, Eqs.~(\ref{eq:R}) and (\ref{eq:GL}) are linear systems of equations (LSE) whose outputs are the desired entries of $\mathbf{C}$. Each $\mathcal{E}$ is independent of the other and can therefore be treated in parallel. On the other hand, the solution of Eq.~(\ref{eq:S}) involves an energy convolution (EC) where a large number of energies is needed for a few of the $\mathcal{C}$ matrix entries. We have set up a parallel data transposition scheme to adapt the distribution of the multi-dimensional $\mathbf{B}$ and $\mathbf{C}$ tensors in memory, depending on the task at hand (LSE or EC). It is schematized in Fig.~\ref{fig2} in the case of electron-electron and electron-phonon interactions.

Concretely, Eqs.~(\ref{eq:R}) and (\ref{eq:GL}) can be solved with the so-called recursive Green's function (RGF) algorithm, which is sequential in essence \cite{rgf}. To simulate larger device structures, we have devised a parallel strategy and implemented it into an open-source, GPU-based library, Serinv. It takes advantage of the Schur complement of BT matrices \cite{serinv} to divide all input/output matrices into multiple partitions. Its principle is summarized in Fig.~\ref{fig3}. The energy convolution in Eq.~(\ref{eq:S}) is performed more efficiently in the time domain after a fast Fourier transform (FFT) \cite{thygesen}.

\begin{figure}[!t]
\center
\includegraphics[width=\linewidth]{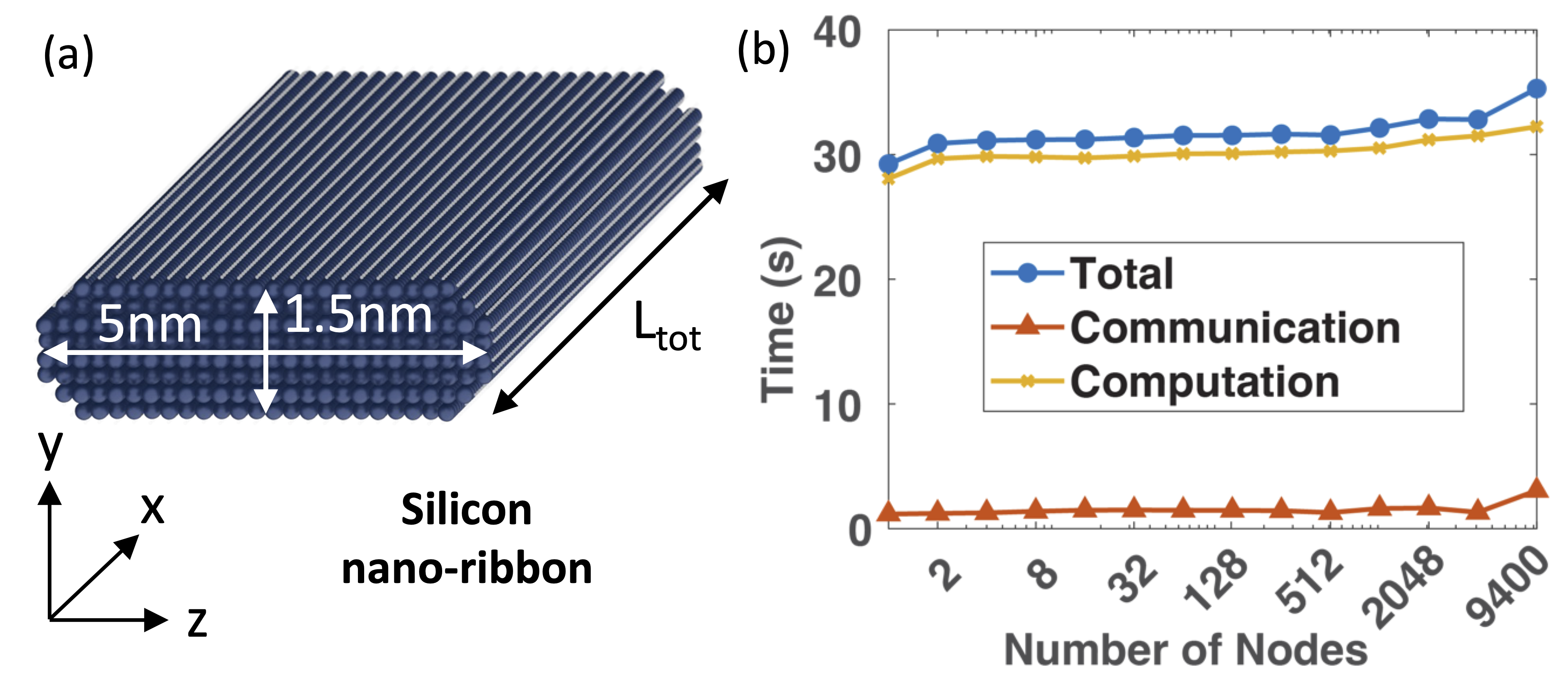}
\caption{(a) Schematic view of a Si nano-ribbon with a height of 1.5 nm, a width of 5 nm, and a varying length $L_{tot}$. The surface of the nano-ribbon is passivated with hydrogen atoms. (b) Weak scaling results for the Si nano-ribbon in (a) with $L_{tot}$=52.1 nm and a total of $N_A$=25,344 atoms, in the presence of electron-electron interactions within the GW approximation, on the Frontier supercomputer. The experiment starts by measuring the execution time on 1 node of the machine with $N_E$=2 energy points, each of them being handled by 4 GPUs. Subsequently, $N_E$ increases proportionally with the number of nodes, reaching 18,800 on 9,400 nodes. The total time is decomposed into its communication and computation components.}
\label{fig4}
\end{figure}

\section{Results}
As an example, we consider a silicon nano-ribbon with cross-section dimensions similar to experiments \cite{intel_2024} (Fig.~\ref{fig4}(a)), including electron-electron interactions (self-consistent GW approximation). Its length is first set to 52.1 nm for a total of $N_A$=25,344 atoms, each of them expressed in a MLWF basis. By leveraging the parallelization approach of Fig.~\ref{fig2} and the distributed NEGF solver of Fig.~\ref{fig3}, we could simulate this device up to the full scale (9,400 nodes, 8 GPUs each) of the Frontier supercomputer at Oak Ridge National Laboratory, reaching a weak scaling parallel efficiency of 80\% when going from 1 to 9,400 nodes (Fig.~\ref{fig4}(b)).

\begin{figure}[!h]
\center
\includegraphics[width=\linewidth]{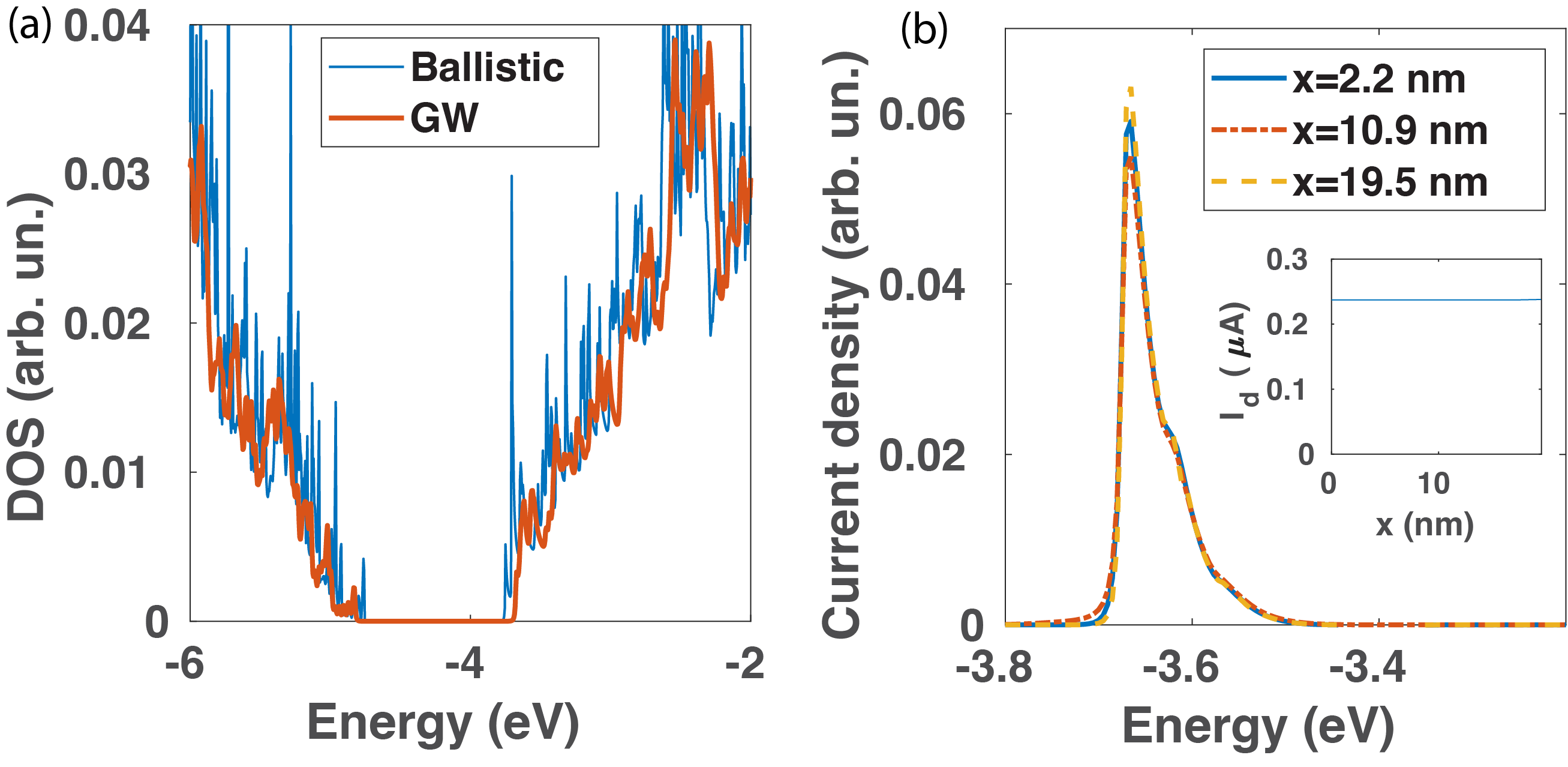}
\caption{Simulation results for the Si nano-ribbon in Fig.~\ref{fig4}(a) with $L_{tot}$=21.7 nm. A linear potential drop is applied with a voltage difference of 0.2 V between both ribbon extremities. (a) Local density-of-states extracted at $x$=0 in the ballistic limit of transport (thin blue line) and in the presence of electron-electron interactions within the GW approximation (thick red curve). (b) Spectral current distribution extracted at $x$=2.2, 10.9, and 19.5 nm. The inset shows the integral of this quantity (electrical current) as a function of $x$, demonstrating current conservation.}
\label{fig5}
\end{figure}

Due to the high computational burden of such calculations, we could not execute a self-consistent Schr\"odinger-Poisson simulation of a transistor with this nano-ribbon as channel material. Instead, we had to reduce its length to 21.7 nm, keeping the cross section the same, and imposed a 10 nm-long linear potential drop with a 5 nm-long flat region on each side of the structure. The potential difference was set to 0.2 V. Selected results are presented in Fig.~\ref{fig5}. From the density-of-states plots in (a), it can be observed that turning on electron-electron interactions (slightly) increases the nano-ribbon band gap, as expected. This limited increase can be explained by the fact that we employ a self-consistent GW scheme, contrary to most DFT packages which implement the G$_0$W$_0$ method. Secondly, the electron concentration is high enough (in the order of 1e16 cm$^{-3}$) to reduce the impact of carrier-carrier scattering on the band gap correction.

The influence of electron-electron interactions on the electronic current distribution is moderate, as can be seen in Fig.~\ref{fig5}(b). While the magnitude of the current's peak and its low-energy tail vary along the nano-ribbon transport direction ($x$), they nevertheless remain very close to each other. Still, it is worth noticing that in spite of these variations, the electronic current remains perfectly conserved from one side of the device to the other, validating the implementation of the model, as also demonstrated in \cite{deuschle}.

\begin{figure*}[!t]
\center
\includegraphics[width=0.95\linewidth]{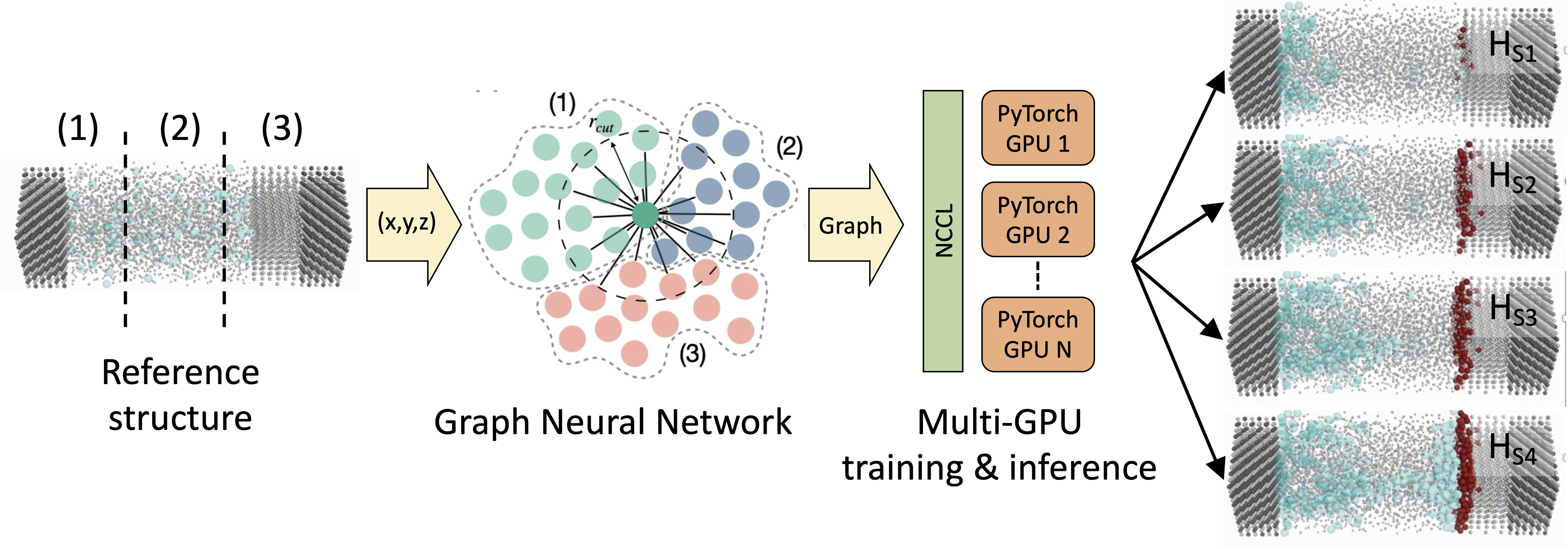}
\vspace{-0.25cm}
\caption{Illustration on how machine learning can be inserted to accelerate \textit{ab-initio} quantum transport simulations by predicting the entries of Hamiltonian matrices $H$ based on training from DFT. A single device structure (left), here a valence change memory (VCM) cell, is divided into several partitions and fed to an equivariant graph neural network (EGNN). Once trained with DFT data from the original VCM on multiple GPUs, this EGNN can accurately predict the Hamiltonian matrix of different VCM configurations with various oxygen vacancy distributions. During prediction, the average error of the $H$ entries does not exceed $\sim$2 meV with respect to DFT \cite{icml}. Note that EGNNs scale with $O(N)$, where $N$ is the number of atoms, and DFT with $O(N^3)$.}
\label{fig6}
\end{figure*}

\begin{figure}[h]
\center
\includegraphics[width=0.85\linewidth]{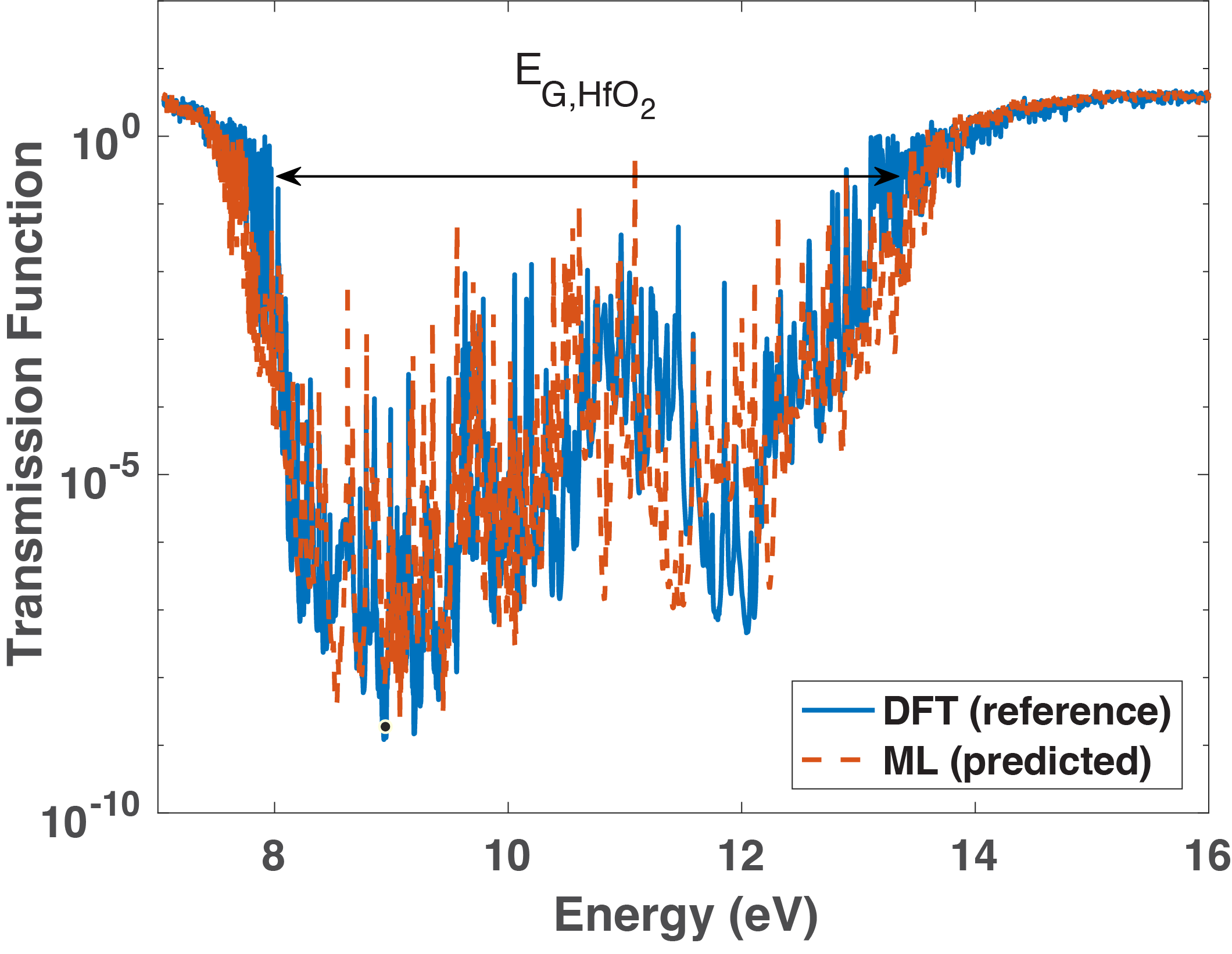}
\caption{Transmission function through a TiN-HfO$_2$-Ti/TiN valence change memory cell as computed from \textit{ab-initio} quantum transport with a DFT Hamiltonian (solid blue curve) and with a Hamiltonian predicted from machine learning after proper training (dashed red curve). The band gap of the HfO$_2$ insulator layer, $E_{G,HfO_2}$, is indicated with a double arrow.}
\label{fig7}
\end{figure}

\section{Outlook: Machine Learning}
Through parallelization and dedicated algorithms, the solution of the NEGF equations can be massively accelerated, which enables the treatment of large atomic systems at unprecedented levels of accuracy. However, the construction of the key ingredient for \textit{ab-initio} QT simulations, the DFT Hamiltonian matrix, remains a challenge, especially in structures made of thousands of atoms. Indeed, DFT scales with $O(N^3)$, $N$ being the number of atoms. Recently, machine-learning techniques based on equivariant graph neural networks (EGNNs) have been proposed to accurately predict the entries of Hamiltonian matrices, after training with DFT data \cite{deeph}. Such approaches scale with $O(N)$, significantly outperforming DFT, but the training cost can be expensive, and most implementations are limited to small domains ($<$150 atoms).

To be of practical relevance, machine-learned methods should allow for the prediction of multiple Hamiltonian matrices from a single training set. Devices with atomic geometries evolving over time are ideal candidates for that. For example, by applying a voltage to a valence change memory (VCM) cell, oxygen vacancies start moving from one electrode to the other, altering the resistance of this component in the process. To simulate the ``current vs. voltage'' characteristics of VCMs, their Hamiltonian matrix must therefore be recomputed at each applied voltage. To speed up such calculations, we have developed an EGNN capable of producing the Hamiltonian matrix of devices containing thousands of atoms and have shown that it can return the desired entries for different oxygen vacancy distributions using one single configuration as training \cite{icml}. Our methodology is displayed in Fig.~\ref{fig6}, whereas the energy resolved transmission function through a VCM cell with a DFT and predicted Hamiltonian is reported in Fig.~\ref{fig7}. Although the average error in the machine-learned Hamiltonian entries ($\sim$2 meV) is comparable to the state-of-the-art for molecules, it is not yet accurate enough to fully reproduce the behavior of the transmission function. 

\section{Conclusion}
In this paper, we outlined a parallelization strategy and numerical algorithm tailored to \textit{ab-initio} quantum transport simulations based on DFT and NEGF. Together, they unlock the possibility to explore close-to-experiment device sizes, including relevant physical effects, e.g., carrier-carrier scattering. By further exploiting machine learning, novel opportunities arise, such as the partial elimination of DFT calculations.


\bibliographystyle{IEEEtran}
\bibliography{sispad_2025}

@article{deuschle,
  title = {{Electron-electron interactions in device simulation via nonequilibrium Green's functions and the GW approximation}},
  author = {Deuschle, Leonard and Cao, Jiang and Ziogas, Alexandros Nikolaos and Winka, Anders and Maeder, Alexander and Vetsch, Nicolas and Luisier, Mathieu},
  journal = {Phys. Rev. B},
  volume = {111},
  pages = {195421},
  numpages = {18},
  year = {2025},
  publisher = {American Physical Society},
  url = {https://link.aps.org/doi/10.1103/PhysRevB.111.195421}
}

@article{taylor,
  title = {Ab initio modeling of quantum transport properties of molecular electronic devices},
  author = {Taylor, Jeremy and Guo, Hong and Wang, Jian},
  journal = {Phys. Rev. B},
  volume = {63},
  issue = {24},
  pages = {245407},
  numpages = {13},
  year = {2001},
  month = {Jun},
  publisher = {American Physical Society},
  url = {https://link.aps.org/doi/10.1103/PhysRevB.63.245407}
}

@article{petersen,
	title = {A hybrid method for the parallel computation of {Green}’s functions},
	volume = {228},
	url = {https://www.sciencedirect.com/science/article/pii/S0021999109001648},
	number = {14},
	journal = {Journal of Computational Physics},
	author = {Petersen, Dan Erik and Li, Song and Stokbro, Kurt and Sørensen, Hans Henrik B. and Hansen, Per Christian and Skelboe, Stig and Darve, Eric},
	year = {2009},
	pages = {5020--5039}
}

@Article{klimeck,
author={Klimeck, Gerhard},
title={Parallelization of the Nanoelectronic Modeling Tool ({NEMO 1-D}) on a Beowulf Cluster},
journal={Journal of Computational Electronics},
year={2002},
month={Jul},
volume={1},
number={1},
pages={75-79},
url={https://doi.org/10.1023/A:1020767811814}
}

@article{darve,
  title={Computing entries of the inverse of a sparse matrix using the {FIND} algorithm},
  author={Li, Song and Ahmed, S and Klimeck, Gerhard and Darve, Eric},
  journal={Journal of Computational Physics},
  volume={227},
  number={22},
  pages={9408--9427},
  year={2008},
  publisher={Elsevier}
}

@inproceedings{ziogas, 
author = {Ziogas, Alexandros Nikolaos and Ben-Nun, Tal and Fern\'{a}ndez, Guillermo Indalecio and Schneider, Timo and Luisier, Mathieu and Hoefler, Torsten}, 
title = {A data-centric approach to extreme-scale ab initio dissipative quantum transport simulations}, 
year = {2019}, 
url = {https://doi.org/10.1145/3295500.3357156}, 
booktitle = {Proceedings of the International Conference for High Performance Computing, Networking, Storage and Analysis}, 
articleno = {1}, 
numpages = {13}, 
location = {Denver, Colorado}, 
series = {SC '19} }

@article{thygesen,
  title = {Conserving {GW} scheme for nonequilibrium quantum transport in molecular contacts},
  author = {Thygesen, Kristian S. and Rubio, Angel},
  journal = {Phys. Rev. B},
  volume = {77},
  pages = {115333},
  numpages = {22},
  year = {2008},
  publisher = {American Physical Society},
  url = {https://link.aps.org/doi/10.1103/PhysRevB.77.115333}
}

@article{cp2k,
  year = {2020},
  publisher = {{AIP} Publishing},
  volume = {152},
  pages = {194103},
  author = {K\"{u}hne \textit{et al.}, T. D.},
  title = {{CP}2K: An Electronic Structure and Molecular Dynamics Software Package - Quickstep: Efficient and Accurate Electronic Structure Calculations},
  journal = {J. Chem. Phys.}
}

@article{wannier,
  title = {Maximally localized generalized {Wannier} functions for composite energy bands},
  author = {Marzari, Nicola and Vanderbilt, David},
  journal = {Phys. Rev. B},
  volume = {56},
  pages = {12847--12865},
  numpages = {0},
  year = {1997},
  publisher = {American Physical Society},
  url = {https://link.aps.org/doi/10.1103/PhysRevB.56.12847}
}

@article{rgf,
  title = {Two-Dimensional Quantum Mechanical Modeling of Nanotransistors},
  author = {Svizhenko, A. and Anantram, M. P. and Govindan, T. R. and Biegel, B. and Venugopal, R.},
  year = {2002},
  journal = {J. Appl. Phys.},
  volume = {91},
  pages = {2343--2354},
  url = {https://doi.org/10.1063/1.1432117}
}

@misc{serinv,
      title={Serinv: A Scalable Library for the Selected Inversion of Block-Tridiagonal with Arrowhead Matrices}, 
      author={Vincent Maillou and Lisa Gaedke-Merzhaeuser and Alexandros Nikolaos Ziogas and Olaf Schenk and Mathieu Luisier},
      year={2025},
      eprint={2503.17528},
      archivePrefix={arXiv},
      primaryClass={cs.DC},
      url={https://arxiv.org/abs/2503.17528}, 
}

@INPROCEEDINGS{intel_2024,
  author={Agrawal \textit{et al.}, A.},
  booktitle={2024 IEEE International Electron Devices Meeting (IEDM)}, 
  title={Silicon RibbonFET {CMOS} at 6nm Gate Length}, 
  year={2024},
  volume={},
  number={},
  pages={1-4},
  url={https://ieeexplore.ieee.org/document/10873367}
}

@article{deeph,
  title = {Deep-learning density functional theory {Hamiltonian} for efficient ab initio electronic-structure calculation},
  volume = {2},
  url = {http://dx.doi.org/10.1038/s43588-022-00265-6},
  number = {6},
  journal = {Nature Computational Science},
  publisher = {Springer Science and Business Media LLC},
  author = {Li,  He and Wang,  Zun and Zou,  Nianlong and Ye,  Meng and Xu,  Runzhang and Gong,  Xiaoxun and Duan,  Wenhui and Xu,  Yong},
  year = {2022},
  pages = {367–377}
}

@misc{icml,
      title={Learning the Electronic {Hamiltonian} of Large Atomic Structures}, 
      author={Chen Hao Xia and Manasa Kaniselvan and Alexandros Nikolaos Ziogas and Marko Mladenović and Rayen Mahjoub and Alexander Maeder and Mathieu Luisier},
      year={2025},
      eprint={2501.19110},
      archivePrefix={arXiv},
      primaryClass={cond-mat.mtrl-sci},
      url={https://arxiv.org/abs/2501.19110}, 
}


\end{document}